\documentclass[showpacs,twocolumn,aps,floatfix,superscriptaddress]{revtex4}
\usepackage{amsmath,amssymb,eucal,graphicx,subfigure}
\usepackage{color} % used for highlighting comments (to remove when doing the resubmission)
\usepackage[colorlinks=true,citecolor=cyan]{hyperref} % citation color
\usepackage{tabu}

\def\av#1{\langle #1\rangle}

\begin{document}

\author{Shandeepa Wickramasinghe}
\affiliation{Clarkson Center for Complex Systems Science ($C^3S^2$), Potsdam, NY 13699, USA}
\affiliation{Department of Mathematics, Clarkson University, Potsdam, NY 13699, USA}

\author{Onyekachukwu Onyerikwu}
\affiliation{Department of Computer Science, Clarkson University, Potsdam, NY 13699, USA}

\author{Jie Sun}
\email{sunj@clarkson.edu}
\affiliation{Clarkson Center for Complex Systems Science ($C^3S^2$), Potsdam, NY 13699, USA}
\affiliation{Department of Mathematics, Clarkson University, Potsdam, NY 13699, USA}
\affiliation{Department of Computer Science, Clarkson University, Potsdam, NY 13699, USA}
\affiliation{Department of Physics, Clarkson University, Potsdam, NY 13699, USA}

\author{Daniel ben-Avraham}
\email{qd00@clarkson.edu}
\affiliation{Clarkson Center for Complex Systems Science ($C^3S^2$), Potsdam, NY 13699, USA}
\affiliation{Department of Mathematics, Clarkson University, Potsdam, NY 13699, USA}
\affiliation{Department of Physics, Clarkson University, Potsdam, NY 13699, USA}

%\today

\title{Modeling spatial social complex networks for dynamical processes
}
%\date{\today}

\begin{abstract}
{The study of social networks --- where people are located, geographically, and how they might be connected to one another --- is a current hot topic of interest, because of its immediate relevance to important applications, from devising efficient immunization techniques for the arrest of epidemics, to the design of better transportation and city planning paradigms, to the understanding of how rumors and opinions spread and take shape over time. We develop a spatial social complex network (SSCN) model that captures not only essential connectivity features of real-life social networks, including a heavy-tailed degree distribution and high clustering, but also the spatial location of individuals, reproducing Zipf's law for the distribution of city populations as well as other observed hallmarks. 
We then simulate  Milgram's Small-World experiment on our SSCN model, obtaining good qualitative agreement with the known results and shedding light on the role played by various network attributes and the strategies used by the players in the game.   This demonstrates the potential of the SSCN model for the simulation and study of the many social processes mentioned above, where both connectivity and geography play a role in the dynamics.}
\end{abstract}

\pacs{89.75.Hc, % Networks and genealogical trees
02.50.-r  % Probability theory Stochastic processes and Statistics
}

\maketitle

\section{Introduction}
Much research has focused in recent years on a wide class of dynamical processes that take place in large human populations, at the scale of cities, whole countries, and even world-wide.  Examples include epidemics spreading and strategies to arrest their spread~\cite{Pastor-Satorras2001,Cohen2003,Belik2011}, the evolution of the electoral map during elections~\cite{Kim2003}, the spreading of rumors~\cite{Kwon2013}, memes~\cite{Hu2014,Gleeson2016} and opinions~\cite{Weng2012}, the migration patterns of banknotes~\cite{Brockmann2006} and human populations~\cite{Lee2014}, and the effects of cities and infra-structure layouts on commerce and productivity~\cite{Bettencourt2013,Schlapfer2014}.  Many of these questions require specific knowledge of individuals' {\em geographical location}  as well as their {\em social contacts} (many infections propagate by direct contact, or physical proximity; we discuss and influence  the opinions of mostly those close to us, etc.).  

In Milgram's Small-World experiment~\cite{milgram}, for example, participants were asked to pass a message (a postcard) to a person in a disclosed address, but only through a chain of social acquaintances: each participant was allowed to pass the message only to a person they know on a first-name basis.  Of 160 messages started in Omaha, Nebraska, 44, or about $28\%$ reached the target in Boston,
Massachusetts, with an average path length of about 5.4 links.  How does the message find its way, let alone in such a short numbers of steps!?

Kleinberg's seminal work~\cite{kleinberg00}, for nodes in a square lattice with random long-range connections, provided a first clue.   This was later extended to  fractal~\cite{roberson06} and anisotropic~\cite{campuzano08} lattices --- still a far cry, however, from the geographical spread and network of connections typical of human society.
Dodds, Muhamad and Watts  conducted a large-scale online experiment that resembles Milgram's original study, highlighting the role of  information beyond just network structure~\cite{dodds03}.
Liben-Nowell {\it et.~al.}~\cite{liben-nowell05} proposed  a spatial social network model with connections derived from an online bloggers community, and studied greedy routing on that model.  Similar studies were conducted for online social networks~\cite{scellato11}  and community structures from mobile phone records~\cite{expert11,onnela11} (see Ref.~\cite{barthelemy11,barthelemy16} for a more comprehensive review). Information on people's location, along with their social contacts, is generally  hard to come by and  often relies on indirect proxies.

In~\cite{Frasco2014} we introduced a stochastic prototype Spatial Social Complex Network (SSCN), relying on just two controllable parameters, that simulates large populations, including the locations and the complex network of contacts between agents.  This
``baseline model" was designed with modest goals in mind: (a)~The population density resembles the light density observed in satellite pictures of earth at night, (b)~the population of ``cities" (defined by percolation clusters~\cite{Rozenfeld2008})
and their rankings follow Zipf's law~\cite{Zipf1949,Cristelli2012}, (c)~the social network of contacts exhibits a scale-free distribution, and
(d)~highly connected nodes tend to be located in denser and larger population areas.   
In addition to meeting these basic goals the SSCN baseline model also yielded good qualitative agreement with census data for the population density as a function of city size, and for the weak super-linear dependence of the cumulative degree of nodes in a city on its total population, as suggested from cell-phone data~\cite{Schlapfer2014}.  Finally, it allowed us to shed some light on the weak deviations~\cite{Rozenfeld2008} from Gibrat's law (that the rate of growth of a city and its fluctuations are proportional to its population size).

Despite these initial successes, the SSCN baseline model fails to mimic real SSCNs in some crucial ways: 
(i)~The complex network of social contacts, while displaying a realistic scale-free degree distribution, is actually a {\em tree}, in contrast with the high degree of clustering observed in social nets and their proxies. (ii)~The network of contacts is built through a redirection mechanism~\cite{KR} which is an adequate description of how individuals might join a social network, but fails to account for the effect of {\em relocations}: every so often a person relocates to a far-away destination, for study, job, or other reasons.  This creates particular correlations in the network of social contacts that are absent in the baseline SSCN.  

In this paper we correct the baseline SSCN deficiencies with some simple adjustments. Connections to spatially closest neighbors are added to mimic the clustering effect in real social networks, and relocations turn out to be crucial in reducing the average path length between nodes.  Simulations of Milgram's Small-World experiment on the revised SSCN model achieve a good qualitative fit with the empirical findings.  This demonstrates the suitability of the model as a substrate for simulations of other dynamic social processes that depend  both on the contacts and the geographical locations of the agents.

\section{The baseline SSCN model}\label{sec:baseline}
We now review the original, or ``baseline" SSCN model,
 established in our previous work~\cite{Frasco2014}.
The model produces a spatially embedded network $G=(V;E;X)$ where $V=\{1,2,\dots,N\}$ is a set of nodes and $E\subset V\times V$ is a set of undirected edges. The spatial embedding of the network is encoded in the set of coordinates $X=\{\mathbf{x}^{(1)},\dots,\mathbf{x}^{(N)}\}$, where for 2D spatial networks (such as in our case) $\mathbf{x}^{(i)}\in\mathbb{R}^2$. A unique feature of the model is that it not only produces the requisite scale-free degree distribution for the edges but it also captures essential spatial features, such as  a Zipf distribution of the populations emerging from the nodes clustering into ``cities"~\cite{Frasco2014}.

Consider first the creation of nodes and edges in the baseline model, defining $V$ and $E$. The starting point is an initial ``seed" network, which in the baseline model consists typically of a single node.
Nodes are added to the network one at a time, each contributing to a new edge, according to a variant of the Krapivsky-Redner (KR) model~\cite{KR} with a single parameter $r\in[0,1]$ --- the {\it redirection probability}. Each time  a new node $i$ joins the network, one of the existing nodes, $j$, is chosen uniformly at random and $i$ is connected to $j$ {\em directly} with probability $1-r$ (creating a new edge $i\leftrightarrow j$); otherwise, with probability $r$, the connection is {\em redirected} to a randomly selected neighbor $j'$ of $j$ (edge $i\leftrightarrow j'$). For large $N$, this leads to a scale-free degree distribution~\cite{kr_remark} 
\begin{equation}\label{eq:gamma}
	P(k)\propto k^{-\gamma},~\mbox{with}~\gamma\approx 1+1/r\,.
\end{equation}

\begin{figure*}[htbp]
\centering
\includegraphics*[width=0.81\textwidth]{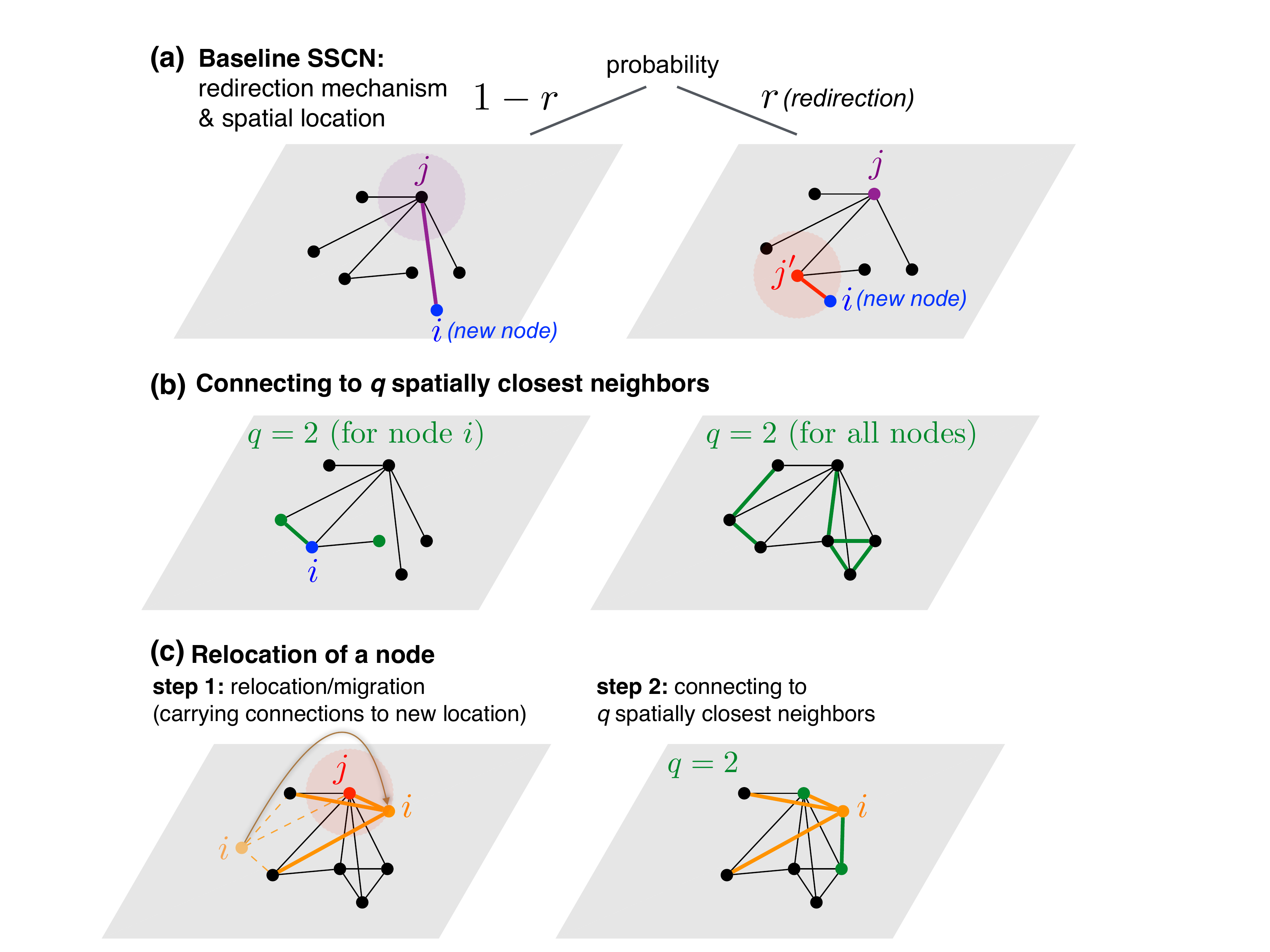}\vspace{-0.05in}
\caption{Growth rules for the baseline~(a) and the revised SSCN model (b \& c).
(a)~A new node $i$ joins the network and connects {\em directly} to a randomly selected node $j$, with probability $1-r$, settling away from $j$ according to the rule of Eq.~(\ref{eq:dist}) (left panel). With probability $r$ the connection is {\em redirected}  to a random
neighbor $j'$ of $j$  and $i$ settles at distance $1$ from $j'$ (right panel). 
(b)~Befriending $q$ closest neighbors (shown for $q=2$). Left panel: Node $i$ needs to add a connection to the nearby node on its left, in order to fulfill the requirement of connections to at least $q$ nearest neighbors.  The new link and $i$'s 2 nearest friends are highlighted in green.  Right: The process is repeated for all nodes in the network until all fulfill the minimum-$q$ requirement.  The new links added to the baseline model are highlighted in green.
(c)~Relocation of node $i$ happens in two stages.  Left: In the first stage $i$ is translated to within distance 1 from a randomly selected node $j$.  All of $i$'s old contacts (broken orange lines) are retained (orange lines).  Right: In the second stage links  are added to ensure connection to at least $q$ new closest neighbors of $i$ (shown for $q=2$). A new link and the 2 closest neighbors are highlighted in green.
}
\label{fig_cartoon}
\end{figure*}

Consider next the placement of the nodes in space, specifying $X$. For a network of $N$ nodes, the baseline model places them within a square box of sides $L=\sqrt{N}$ (with periodic boundary conditions).
The initial seed node is placed at the origin, $\mathbf{x}_1=\mathbf{0}$, and
the location of  subsequent nodes $i$ depends on whether it connects to node $j$ directly or to a neighbor $j'$, by redirection. If $i$ joins directly, it is placed at $(s,\theta)$ from $j$ (using polar coordinates), where the angle $0<\theta\leq 2\pi$ is chosen randomly from the uniform distribution, and $s$ is picked randomly from the distribution
\begin{equation}\label{eq:dist}
p(s)=
\begin{cases}
\frac{1}{\ln(s_{\max})} s^{-1}, &1<s<s_{\max};\\
0,&\mbox{otherwise,}
\end{cases}
\end{equation}
where $s_{\max}=\sqrt{2}L$ is the maximum possible distance between any two points within the bounding square.
In the case of redirection, when node $i$ joins to $j'$, then we simply place $i$ at distance $1$ from $j'$ at a random angle $\theta$.  The growth algorithm is illustrated in Fig.~\ref{fig_cartoon}(a).

%:     FIGURE 1:  Illustration of Model's growth rules
%%%% END OF FIGURE 1 %%%%%

While the above growth rules were ultimately selected to best achieve the baseline model's goals, they do make some intuitive sense as well.  The redirection mechanism introduces a ``rich-get-richer" bias in that redirection favors the random selection of nodes $j'$ of a higher degree.  This accounts for the emergence of the scale-free degree distribution.
In addition, the connection and placement rules capture some basic ways of life:
A person $i$ joins an existing social net when they are born.  There is no choice in this matter and the social connection(s) established in this case is random (direct connection to node $j$).  Eventually $i$ leaves home and settles at some distant location.  The distribution of the distance  to $i$'s new home, inversely proportional to the distance $s$, is motivated by Kleinberg's ``magical" condition for navigability~\cite{kleinberg00}.  The other possibility is that $i$'s most meaningful social connection happens through redirection ($i$ is referred  to a workplace or school, etc.) and in that case it makes sense to settle nearby to the new contact (at distance 1 --- the minimal distance in our distance distribution).

The  growth rules of the baseline SSCN model seem however too simplistic in that they account for a bare minimum of social connections: the connections to one's birth place are represented by a single link, as are also the connections to people in a referred (redirected) situation.  While the sparsity of connections can be justified on the grounds that the model is a scaled-down version of real life (fewer nodes, or people, so fewer contacts per person), there is no getting around the fact that the baseline model network of connections is a {\em tree}, in contrast with real-life social nets, where {\em clustering} is large (your friends have a higher than average probability to be friends among themselves).  Another important effect is that of {\em relocations}: occasionally people move to a different place, sometimes more than once, over the course of their lives.  When people relocate they maintain friendship with some acquaintances in their place of origin, and form friendships with their new neighbors.  Thus relocations have a profound effect on the network of social contacts.  In the next section, we describe a new version of the baseline SSCN that fixes these shortcomings.

\section{A revised SSCN model}

For the present simulations we use a redirection probability $r=0.8$, same as for the baseline model.  This leads to
a degree exponent $\gamma\approx 2.3$ which is typical of large-scale social networks~\cite{BA2003,Newman2003}.  In addition to the significant changes that we made to the model's connectivity, we  made some minor changes to the boundary conditions and to the initial seed, and we describe these first.

\medskip
\noindent
%:  Free boundary condition
{\bf Free boundary condition:}

In the baseline model we used a bounding box of side $L=\sqrt{N}$ and periodic boundary conditions.  For the present work we adopt a boundary-free approach.  Simply, the first node is placed at the origin and each subsequent node is placed in the same fashion as for the baseline model, but without regard to the bounding box.  That is, the nodes are allowed to spread as far as the simulation takes them.  Our simulations show that even with this free boundary condition the radius of gyration scales quite accurately as  $\sqrt{N}$, so that the average population density per unit area remains constant even as the model is scaled up.

\medskip
\noindent
%:  Initial seed
{\bf Initial seed:}

\begin{figure}[htbp]
\centering
\includegraphics*[width=0.47\textwidth]{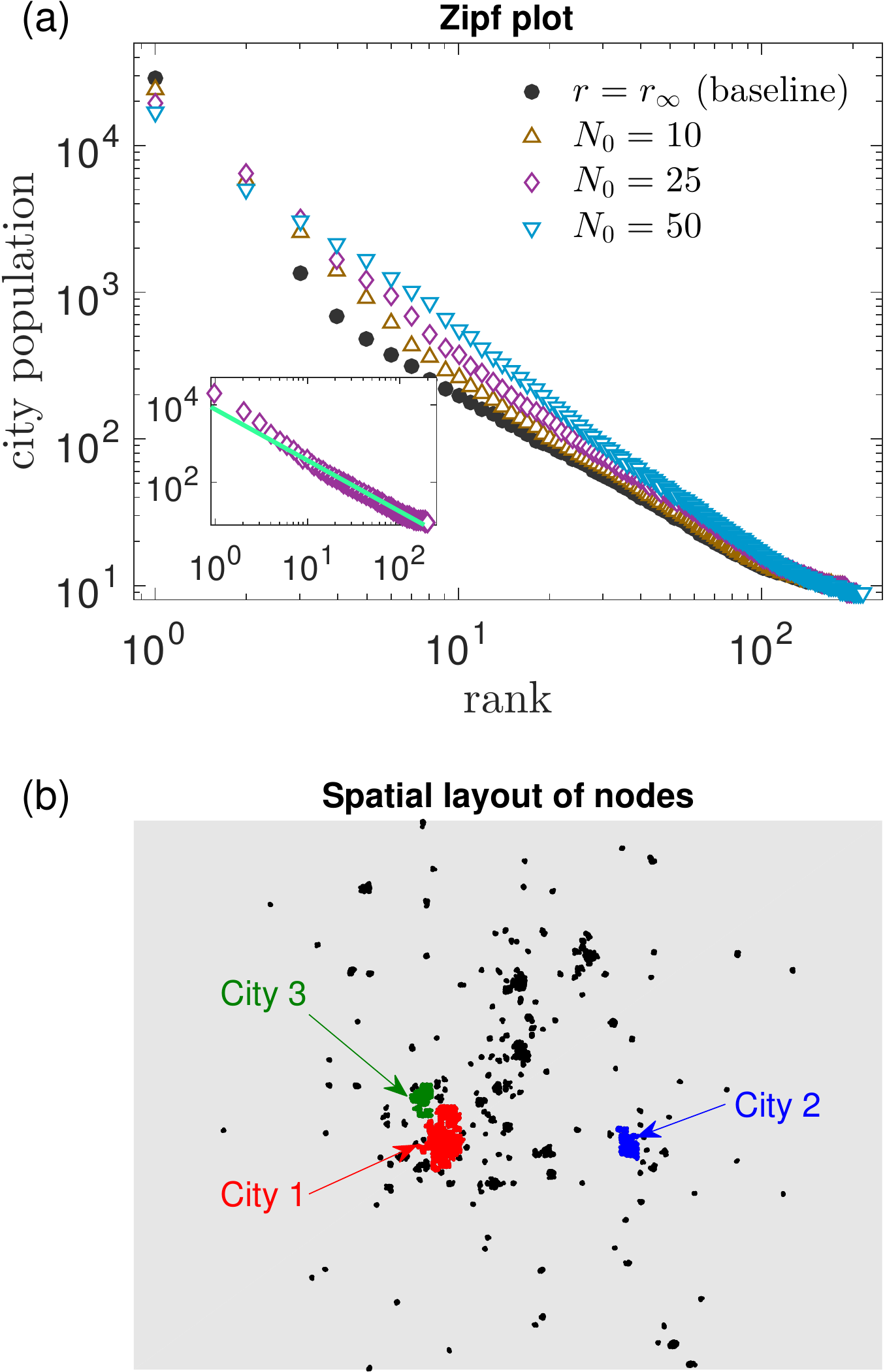}\vspace{-0.05in}
\caption{(a)~Effect of $N_0$ on the distribution of city sizes by rank, on a log-log scale. The inset highlights the case
of $N_0=25$ that we use for our simulations.  The fitted straight line has slope $\approx-1.32$.
(b)~Spatial layout of a network of $N=51200$ nodes generated with $N_0=25$. For visual clarity, we divide the spatial domain into $200$-by-$200$ equal-size square boxes, and  show only the nodes in boxes with a population exceeding the average (per non-empty box). The three largest ``cities" are color-coded in red (pop.~$15,072$), blue (pop.~$5,567$), and green (pop.~$3,743$). The cities were identified by the spatial {\it City Clustering Algorithm}  introduced in~\cite{Rozenfeld2008} and used in~\cite{Frasco2014}.
}
\label{fig_mega}
\end{figure}

Starting with a single-node seed, as in the baseline model, tends to produce a few ``megacities" --- cities that are disproportionately larger than predicted by the Zipf distribution~\cite{Zipf1949,Cristelli2012}.  In~\cite{Frasco2014} we showed how the problem might be overcome by starting with seeds consisting of several nodes.   Here we employ a single-node seed, but let the redirection probability  vary with the number of nodes $i$ added thereafter:
\begin{equation}\label{eq:ri}
	r_i = (1-e^{-(i-1)/N_0})r_{\infty}\,.
\end{equation}
The probability $r_i$ converges rapidly to $r_{\infty}$ (we pick $r_{\infty}=0.8$), and the parameter $N_0$ controls the pace of the convergence. Thus, for $N_0\ll N$ the varying $r_i$ affects mainly the first $\sim N_0$ nodes, but not the large-scale structure of the network.  On the other hand, the fact that $r_i\approx0$ for the first few nodes reduces their capacity to attract further connections, thereby alleviating the problem of megacities.  The effect of $N_0$ on the distribution of city sizes is shown in Fig.~\ref{fig_mega}(a).  In Fig.~\ref{fig_mega}(b) we show the spatial layout of a typical network produced with $N_0=25$, highlighting in color the first three largest cities.  This very same configuration is used for the studies of connectivity and for the simulations of Milgram's Small-World experiment reported below.

%:     FIGURE 2:  Layout of model and  Zipf dist.
%%%% END OF FIGURE 2 %%%%%

\medskip
\noindent
%:  Closest neighbors and clustering
{\bf Closest neighbors and clustering:}

We now come to the more serious revisions of the baseline SSCN model.  A big issue is that the baseline model's network of social contacts is a tree.  This means that the probability for two of your friends to be friends amongst themselves is {\em zero}, while in real life that probability is in fact much higher than the average density of links, an effect best captured by the concept of {\em clustering}~\cite{Watts98,Boccaletti2006}.  

To fix the problem of clustering in the baseline model, we now require that each node be connected
to at least $q$ of its {\em geographically} closest neighbors, mimicking the fact that one indeed tends to befriend ``next-door" neighbors.  New edges are added in at the end of the growth process. The addition of new edges is illustrated in Fig.~\ref{fig_cartoon}(b). Note that the baseline model corresponds to the special case of $q=0$.

\begin{figure}[htbp]
\centering
\includegraphics*[width=0.47\textwidth]{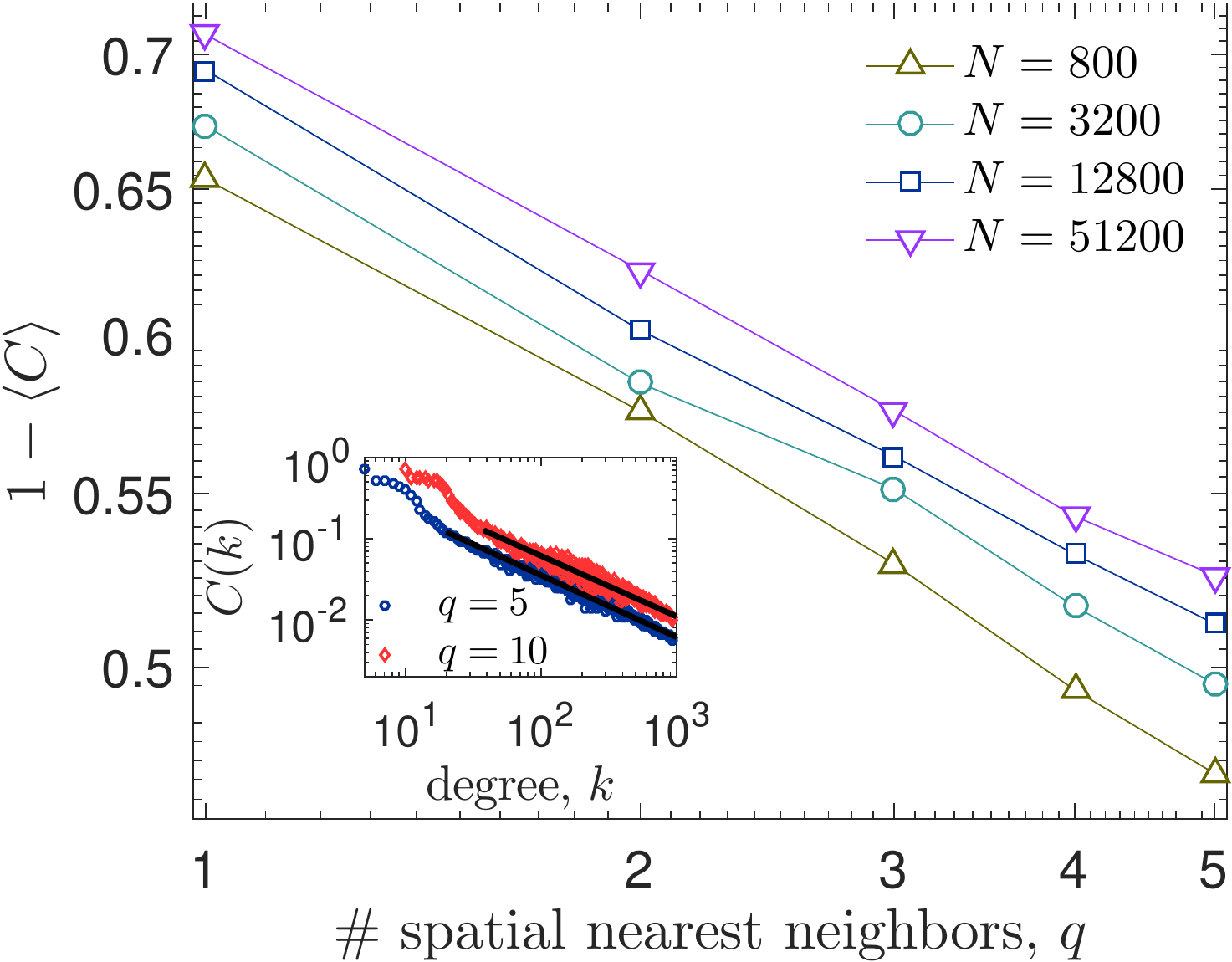}\vspace{-0.05in}
\caption{ 
Dependence of the average clustering coefficient $\langle C\rangle$ on $q$, for networks of size $N=800$, $3,200$, $12,800$, and $51,200$ (from bottom  to top). The slope of the curves in this log-log plot is roughly $-0.2$. Inset: Clustering coefficient $C(k)$ as a function of node degree $k$ for networks of size $N=51,200$, with $q=5$ and $q=10$.  The fitted straight lines have slope $\approx-0.75$.
Each data point in the figures is the result of an average over $20$ independent network generations.
}
\label{fig_clustering}
\end{figure}

In Fig.~\ref{fig_clustering} we plot the  clustering coefficient of the network, $\av{C}$, as a function of $q$.
We see that $\av{C}$ is quite large, and in line with real-life networks, already for $q=1$.  $\av{C}$ grows with $q$ (and decreases with the network size $N$) according to the empirical relation $1 - \langle C\rangle \propto \log(N)q^{-0.2}$.  The inset of the figure shows the dependence of the clustering coefficient of individual nodes upon their degree $k$.  The emergent relation $C(k)\sim k^{-x}$ ($x\approx-0.75$)  is also typical of many real-life networks~\cite{Boccaletti2006}.

%:     FIGURE 3: Clustering
%%%% END OF FIGURE 3 %%%%%

\medskip
\noindent
%:  Relocations
{\bf Relocations:}

The growth rules of the baseline model, even with the added rule for connecting $q$ closest neighbors, still fail to account for the very important effect of relocations.  Every so often a person relocates to a new place, changing jobs or pursuing education, following marriage, etc.  When a person relocates they retain many of their friendships at their place of origin, and form new friendships at their new location.  This has a profound effect on the connectivity of the social network, as we shall see below.  For now, however, we just describe the way to incorporate relocations in the revised SSCN model.

To relocate a single node $i$ we first pick two nodes $i$ and $j$ at random and move node $i$ to within distance $s=1$ from node $j$, and at a random angle $\theta$ from $j$, while retaining all of $i$'s connections.  In the second stage, we examine the new environs of node $i$ and add the necessary connections to enforce the minimum $q$ closest neighbors rule.  Note that the first stage  entails merely changing ${\bf x}^{(i)}$, but not its contacts.  The second stage ensures that agent $i$ not only keeps its old social connections, but also makes new acquaintances in the new place.  The process of relocation is illustrated in Fig.~\ref{fig_cartoon}(c).

The random choice of the relocating node $i$ and the target node (or location) $j$  is motivated by the ``gravity model" for human mobility~\cite{bharti08}.  It basically assumes that any individual $i$ is as likely to relocate as any other, and that relocating to any particular place (near ${\bf x}^{(j)}$) is more probable the more populated that place is.  

In the following section, we study the effect of migrating a fraction $\varepsilon$ of the $N$ nodes in the system.
A single relocation affects the degree of the relocating node $i$ in the same way as adding $q$ closest neighbors. (But note that $i$ undergoes two such updates.)
Thus, the combined effect of connecting $q$ closest neighbors and migrating a fraction $\varepsilon$ on the degree distribution is similar to that of connecting $q'=q(1+\varepsilon)$ neighbors {\em without} migration.  On the other hand, relocations  have a dramatic effect on the {\em pattern} of connections and on navigation of the social network and they should not be neglected.

%%%%%%%%%%
\section{Connectivity and Milgram's Small-World Experiment}

We now turn to the main question of how well the social network is connected and what we can learn from simulations of Milgram's  Small-World experiment.  For concreteness, we study the typical SSCN configuration 
shown in Fig.~\ref{fig_mega}(b) and focus on the connectivity between individuals in the largest  and second-largest cities in the figure (population $15,072$
and $5,567$, respectively).  The two cities happen to be about $190$ units of length away from one another, which compares nicely with $s_{\rm max}=\sqrt{N}\approx226$ and with the actual span of the ``country".

\subsection{Shortest Paths}

Consider first the {\em shortest paths} in the network. Shortest paths can be found very efficiently, for example by the Breadth-First Search (BFS) algorithm. The problem is that efficient algorithms such as the BFS require {\em global} knowledge of the whole network of contacts (or the full adjacency matrix).  This type of information is clearly not available to any one person, so the mere existence of shortest paths cannot explain the results in Milgram's Small-World experiment. Nevertheless, shortest paths constitute a useful ``benchmark" to which one can compare various decentralized algorithms.

%:     FIGURE 4:  Shortest paths
\begin{figure*}[htbp]
\centering
\includegraphics*[width=0.65\textwidth]{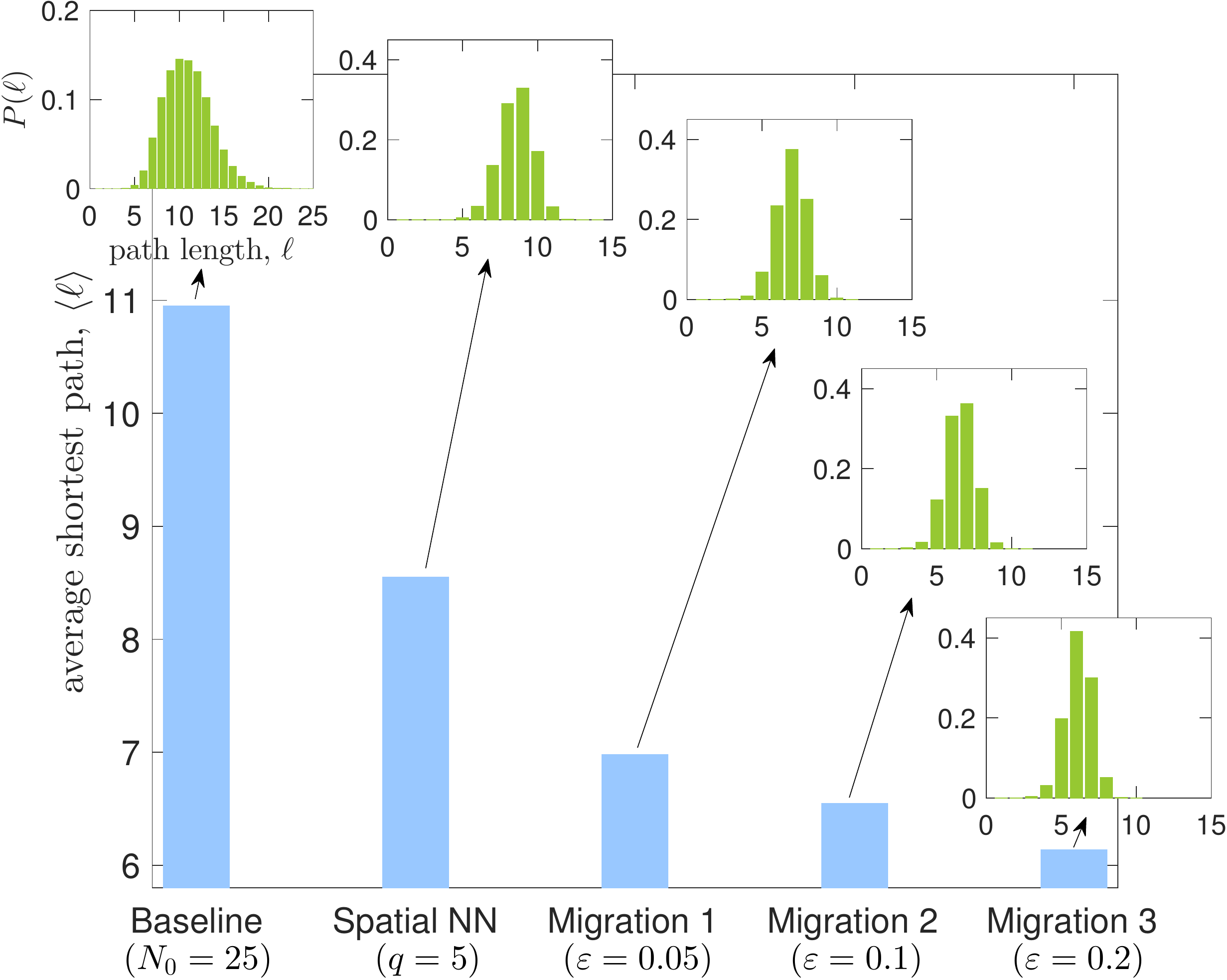}\vspace{-0.05in}
\caption{Statistics of shortest path length between all node pairs $(i,j)$ where $i$ and $j$ belong to cities 1 and 2 as shown in Fig.~\ref{fig_mega}(b), indicating a decrease of the average path length as additional features are introduced into the model, as well as narrowing of their distribution (inset histograms).
}
\label{fig_migration_SP}
\end{figure*}
%%%% END OF FIGURE 4 %%%%%

Since the SSCN network of social contacts consists of only one connected component (even in the baseline model) there exists a shortest path of links between any two nodes.  We explore first how shortest paths evolve as one adds connectivity to the baseline model, first by connecting $q=5$ closest neighbors, then by migrating increasing fractions $\varepsilon=0.05$, $0.1$, and $0.2$ of the nodes.  

Our results for the shortest paths between nodes~$i$ in City~1 and nodes~$j$ in City~2 are summarized in Fig.~\ref{fig_migration_SP}.  For the baseline model, the shortest paths between nodes in the two cities follow a bell-shaped distribution and average to just under 11 links.  Adding connections to 5-closest neighbors reduces the shortest paths average length to about 8.5.  This change is actually less impressive than one would expect:
For a random $N$-nodes network of average degree $\av{k}$ the average shortest path scales as $\log_{\av{k}}N$.
The baseline model has $\av{k}=2$ (it being a tree) and adding  5 closest neighbors increases $\av{k}$ to nearly 7.
Thus the $8.5$ average looks  long compared to the average expected for random nets, of $11\log_7 2\approx3.9$.  The reason is, of course, that the added links are {\em not} random, and --- while important in accounting for the common phenomenon of ``next-door" friends --- they do not create efficient shortcuts.  The situation is quite opposite for relocations: Migrating a mere $0.05$ fraction of the nodes results in an additional shortening of the average path lengths to about 7, a dramatic change for the tiny increase in $\av{k}$, from $7$ to $7.25$.  Increasing the migration rate results in further reduction of the average path lengths, but the most dramatic change is that seen between no relocations at all and a tiny fraction of relocations.  In that respect relocations seem to play a similar role to that of random long-range connections in  the Watts and Strogatz Small-World networks~\cite{Watts98}.  Finally, the insets in the figure show the distribution of path lengths for each successive change.  The narrowing of these distributions can be traced to the homogeneization of the degree distribution as more links are added in.

\medskip
\subsection{Greedy Paths}

Consider now Milgram's Small-World experiment~\cite{milgram}.  Participants in the experiment have access only to {\em local}
information:  You know who your friends are and where they live, etc., but have little information about their friends
down the line.  The puzzle is how the message finds its way, under these circumstances, let alone in a short number of steps.  Local, or {\em decentralized} algorithms for passing the message may be quite involved and we shall test a few scenarios.  For now however, we stick to the simplest {\em greedy algorithm:}
\begin{quote}
{\em Pass the message to the contact that is geographically closest to the target (provided that it is closer than yourself).}
\end{quote}
Kleinberg~\cite{kleinberg00} had shown that, for his Small-World lattice, no other decentralized algorithm can obtain paths that scale more favorably with the population $N$ than the greedy algorithm.  In other words, greedy paths give us a good idea of how well any other decentralized method might perform (at least functionally in $N$).

The proviso that each subsequent node is {\em closer} to the target is important:  On the one hand it guarantees convergence; on the other hand, it means that the message might get stuck, when there is not a single contact that is closer to the target than oneself.  In such a case  there is no greedy path between the source and the target.  When a greedy path exists, we say that the source and target are {\em greedily connected}.  Greedy connectivity was explored
for some benchmark networks (but not for SSCN models) in~\cite{Sun2010}.  Some of the more important properties of greedy connectivity are:
\begin{itemize}\vspace{-0.1in}
\item{Nodes that are connected in the usual sense might not be greedily connected (but not the other way around).}\vspace{-0.1in}
\item{Greedy paths are never shorter than shortest paths.}\vspace{-0.1in}
\item{Greedy connectivity is not transitive: If $u$ is greedily connected to $v$ and $v$ is greedily connected to $w$, it is {\em not} necessarily the case that $u$ is greedily connected to $w$.}\vspace{-0.1in}
\item{Greedy connectivity is  not symmetric: there might be a greedy path from $u$ to $v$ but no greedy path from $v$ to $u$.}\vspace{-0.1in}
\end{itemize}

\begin{figure*}[htbp]
\centering
\includegraphics*[width=0.65\textwidth]{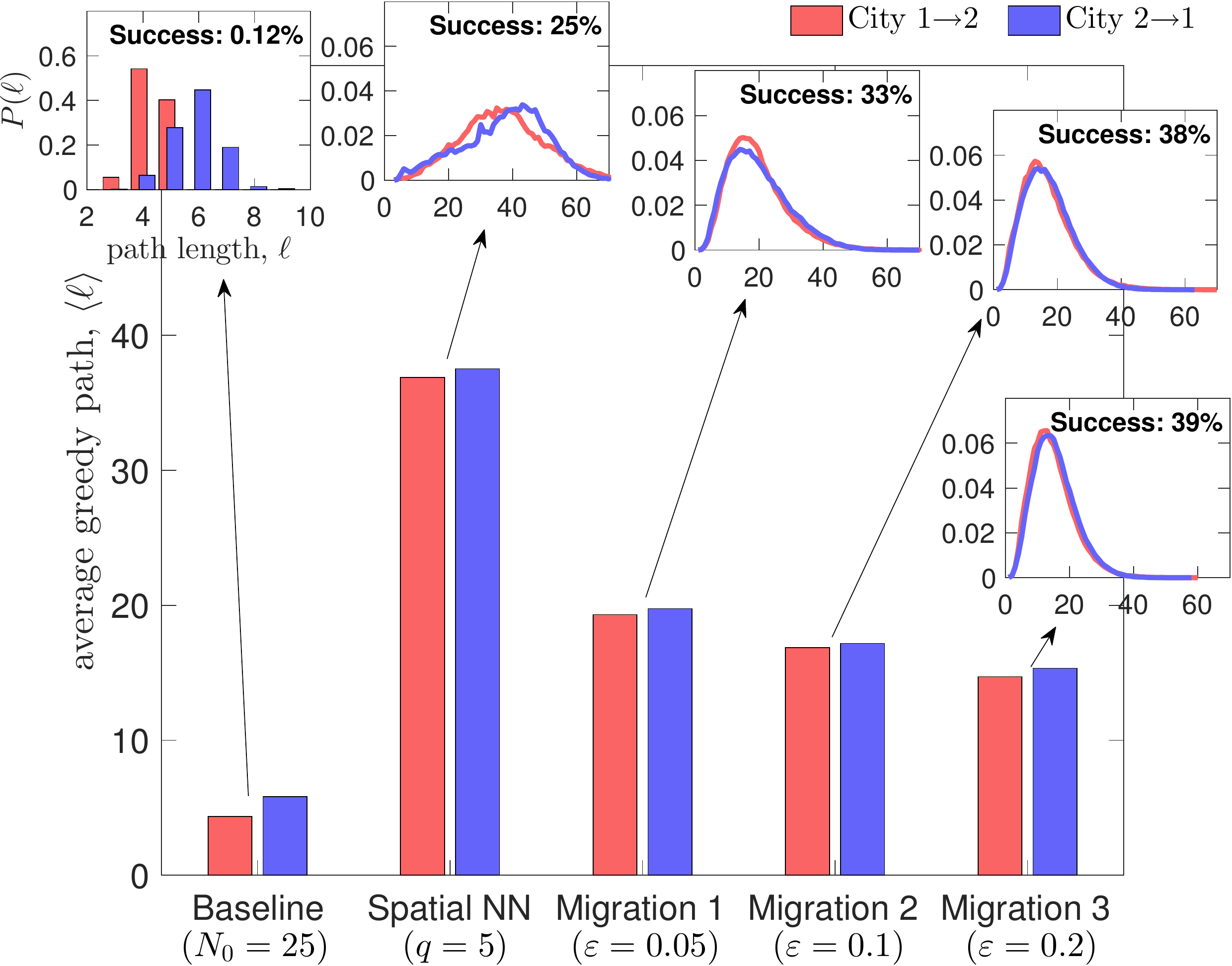}\vspace{-0.05in}
\caption{Statistics of greedy path length obtained by randomly sampling $5\times10^5$ node pairs $(i,j)$ where $i$ and $j$ belong to the cities 1 and 2, see Fig.~\ref{fig_mega}(b). In the baseline model there are very few short, greedy paths. Connecting to closest neighbors increases the success rate significantly, but the paths found are quite longer. Even a tiny percentage of relocations not only further increases the success rate, but also reduces the greedy path length significantly.
}
\label{fig_migration_GP}
\end{figure*}

We have selected $500,000$ random pairs of nodes $(i,j)$, with $i\in{\rm City\ }1$ and $j\in{\rm City\ }2$, and then searched for greedy paths from $i$ to $j$, and from $j$ to $i$.  The results are summarized in
Fig.~\ref{fig_migration_GP}.

%:     FIGURE 5:  Greedy paths
%%%% END OF FIGURE 5 %%%%%

The average greedy path length for the baseline model, of about 7 links, is pleasingly short, however, only $0.12\%$ of the pairs are greedily connected.  Adding connections to $q=5$ closest neighbors dramatically increases the greedy connectivity, to about $25\%$ of the pairs, but the average greedy path lengthens
to about 39 links.  These results can be understood as follows.  In the baseline model the network of contacts is a tree and there is a {\em unique} path between any pair of nodes. (This path is also the shortest path.) Since the spatial connections are lain at a random angle $\theta$, the probability that an $\ell$-links path from $i$ to $j$ be also a greedy path, is
$(1/2)^{\ell}$.  Thus the typical shortest paths, of average length $\av{\ell}=11$, are greedy paths with probability $(1/2)^{11}\approx0.05\%$, in general agreement
with the observed result. Connecting $q$ closest neighbors makes for multiple paths between pairs of nodes. The probability that a greedy search might have to be abandoned at any particular step is roughly
$(1/2)^q$ (assuming that the closest neighbors are randomly distributed, and neglecting the underlying baseline tree).  For $q=5$, the probability of the typical greedy paths (of length 39) making it through is therefore $(1-(1/2)^5)^{39}\approx29\%$, quite in line with the observed results. Despite the dramatic increase in the success
rate for greedy searches, the typical path length is too large to explain the observations in Milgram's Small-World experiment.

Migrating even a small fraction $\varepsilon=0.05$ of the nodes further increases the success rate, to about $33\%$, but more importantly, it slashes the typical greedy path length by a factor of 2.  Migrating larger fractions of the population achieves only modest improvements. Once again, the role of relocations seems analogous to that of random long-range connections in  Watts and Strogatz Small-World networks~\cite{Watts98}.  Nevertheless, the typical greedy path lengths, of about $15$,
even for $\varepsilon=0.2$ migrations, still seems too long to account for Milgram's results.  Our SSCN model suggests that the difference is due largely to clever strategies adopted by participants in the experiment --- people act more cleverly than the simple-minded greedy algorithm --- and partly due to the effect of {\em attrition}: the finite probability to drop the search at any particular step selects for shorter paths.  We turn to these issues next.

\subsection{Complex Strategies and Attrition}
The greedy path algorithm cannot by itself explain the results from  Milgram's Small-World experiment and we are led to consider more complex strategies.
A possible strategy is to prefer friends that live closer to the target  to some extent, but give also some weight to friends that are exceptionally well-connected (since they might be more likely to make a better choice than ourselves).  The following algorithm captures the gist of this idea.

Suppose that node $i$ currently holds the message that is destined for the (disclosed) target $t$.  Node $i$ assigns a score $S_{j}$
to each of his $k_i$ acquaintances ($j=1,2,\dots,k_i$):
  \begin{equation}
 S_{j}=\lambda\frac{s_{i}}{s_{j}}+(1-\lambda)\frac{k_{j}}{k_i}\,.
 \end{equation}
 Here $s_i$ and $s_{j}$ are the geographical distances between $i$ and $t$ and $j$ and $t$, and $k_i$ and $k_{j}$ are the degrees of node $i$
 and of its $j$-th contact, respectively.  In other words, agent $i$ scores his acquaintances relative to himself (his own score is $S_i=1$), assigning higher value to friends that are closer to $t$ than himself, and that are better connected than himself.  The parameter $\lambda\in[0,1]$ controls the relative importance of each attribute.  With the scores at hand the strategy proceeds exactly as in the greedy algorithm, but with the aim of maximizing $S_j$ (rather than minimizing the distance):

\begin{quote}
{\em Pass the message to the contact that has the largest score (provided that its score is larger than 1).}
\end{quote}

Kleinberg's greedy algorithm corresponds to the special case of $\lambda=1$.  For any other $0<\lambda<1$ the strategy still guarantees convergence to the target (if a path is available), since the distance from $t$ to itself is zero, so that the score of $t$ is infinite and overwhelms all other considerations.  (The case of $\lambda=0$ is problematic, for the message may then fail to reach the target, even when $t$ is a contact of $i$, and we therefore require $\lambda>0$.)  The search for a path to $t$ is aborted when the proviso that $S_j>1$ is not fulfilled.  In addition, for $\lambda<1$ the path may revisit a previously touched node, creating a closed loop.  The search is, of course, abandoned in such cases as well.

\begin{figure}[htbp]
\centering
\includegraphics*[width=0.47\textwidth]{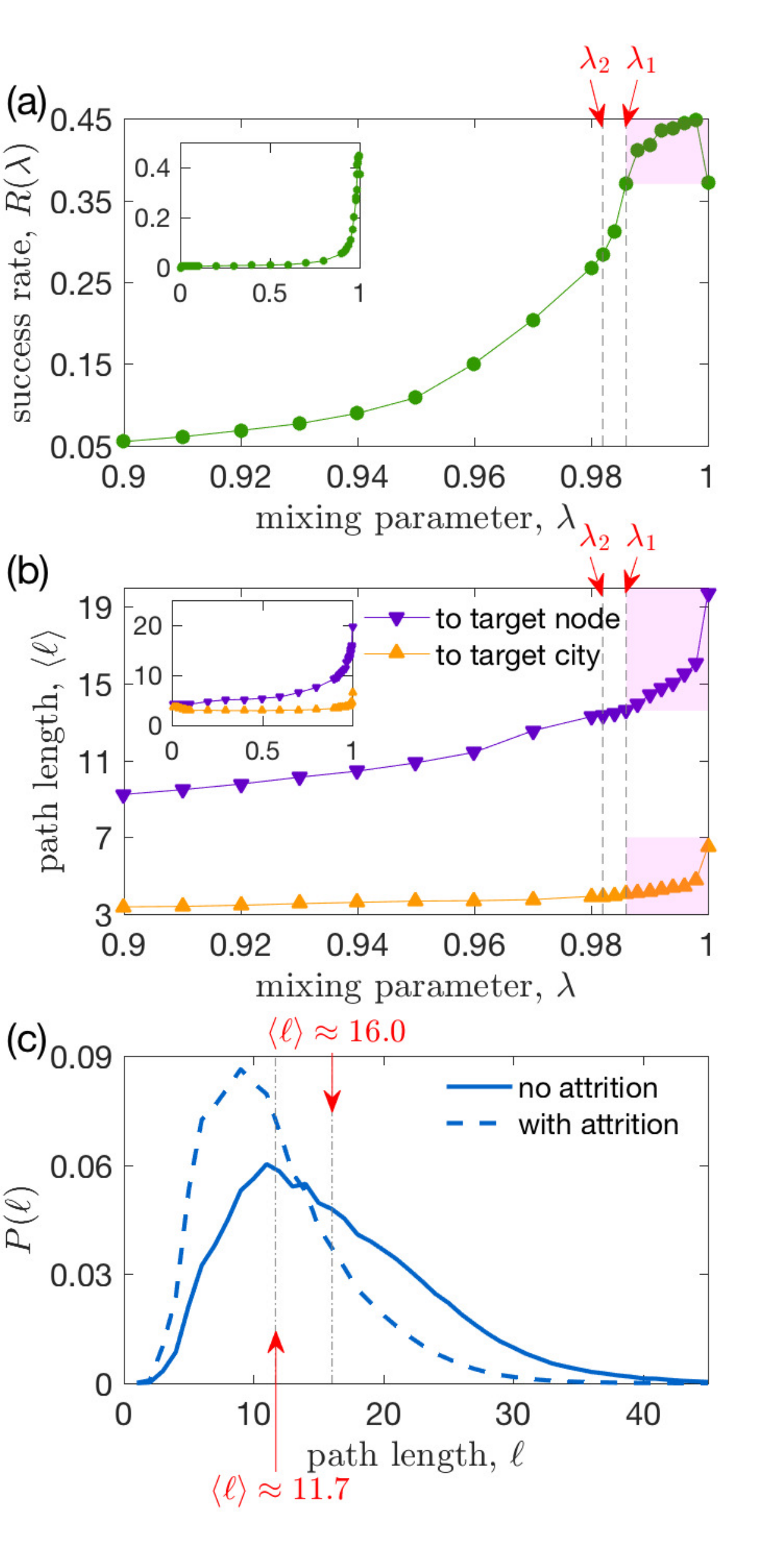}\vspace{-0.2in}
\caption{Decentralized paths found with the mixed greedy strategy.
(a)~Fraction of completed searches $R(\lambda)$ in the range $0.9\leq\lambda\leq1$.  The mixed strategy beats the pure greedy algorithm in the pink-shaded region, $\lambda_1<\lambda<1$.  At $\lambda=\lambda_2$ the success rate of the mixed strategy matches the $28\%$ rate reported in Milgram's work~\cite{milgram}.  Inset: $R(\lambda)$ for the whole range of $0\leq\lambda\leq1$.
(b)~Average path length from points in City~2 to City~1 (top, purple curve) and average number of links to reach
City~1 (bottom, orange curve) in the range $0.9\leq\lambda\leq1$. Inset: Same, for the full range of $0\leq\lambda\leq1$.
(c)~The effect of incidental attrition: Shown is the distribution of path lengths, $P(\ell)$ for $\lambda=0.998$ (solid line) along with $(0.9)^{\ell}P(\ell)$, accounting for $0.1$ probability of incidental dropout (broken line).  The overall success rate reduces from $45\%$ to $11\%$ and the conditional average path length $\av{\ell}$ decreases from 16 to 11.7.  (Both curves are normalized in the figure, to highlight the change in shape that results from incidental attrition.)
}
\label{fig_mixed_GP}
\end{figure}

Fig.~\ref{fig_mixed_GP} summarizes the results of this mixed strategy, as applied to the case of $q=5$ closest neighbors and $\varepsilon=0.05$ fraction of relocations.  For clarity, we include only the results for searches from City~2 to City~1  (the small differences found for the reverse direction are discussed in the next subsection).  Panel~(a) shows the fraction of pairs, $R(\lambda)$, that are successfully connected.  The overall trend, shown in the inset, is of a rapid decay to zero as $\lambda$ decreases. For $\lambda$ close to 1, however, there is first an increase, from $R(1)\approx0.37$ to a maximum of $0.45$ success rate for $\lambda\approx0.998$.  At the same time, the average path length (Fig.~\ref{fig_mixed_GP}(b)) decreases from $\av{\ell}=19.7$ at $\lambda=1$ to $\av{\ell}=16.0$ at $\lambda=0.998$.  There is in fact a whole range of $\lambda_1<\lambda<1$ for which the mixed strategy performs better (higher success rate {\em and} shorter paths) than the pure greedy algorithm of $\lambda=1$.  At $\lambda_1\approx0.986$, for example, the success rate is as good as for $\lambda=1$, but the average path length is slashed by nearly 5 links.

%:     FIGURE 6:  Complex strategies
%%%% END OF FIGURE 6 %%%%%

As $\lambda$ decreases beyond $\lambda_1$ it becomes harder to judge the success of the mixed strategy:  On the one hand there is the attractive effect of decreasing $\av{\ell}$, on the other hand fewer and fewer pairs remain connected.  One way out of this conundrum is to select the point for which $R$ matches the reported success rate of Milgram's Small-World experiment, of roughly $28\%$.  This occurs for $\lambda_2\approx0.982$, where $\av{\ell}$ is reduced to nearly $13.4$ links.

An important conclusion is that geographical proximity is the largest factor in finding decentralized paths,
as evident from the large values of $\lambda$ that are optimal in our mixed strategy.  This understanding is also in line with the findings of Liben-Nowell et al.~\cite{liben-nowell05}. 
Our mixed strategy shows that one can do  better than geography alone  (the case of $\lambda=1$), yet not as well as reported by Milgram.  The reason is that our mixed strategy fails to incorporate much of the intuition and social cleverness that are second-nature to people.  In Milgram's experiment, for example, the target's occupation (stockbroker) was disclosed in addition to  name and address.  The name holds  clues to the target's gender and ethnicity, and the address might hint at social status.  None of this information is accounted for in our naive approach. 

A more realistic approach would probably still rely  mostly on geography, at least until the message reaches the target's city.  Once inside the city the additional clues of occupation, gender, ethnicity, social status, etc.,  provide effective means for finding shorter paths (e.g., the stockbrokers in Boston tend to know one another).  Indeed, subject reports in Milgram-like experiments strongly support this idea~\cite{dodds03}. The average path to the target's city in our simulations is significantly shorter than the total path (Fig.~\ref{fig_mixed_GP}(b)).  At $\lambda_2=0.982$ (where we reproduce Milgram's success rate of $28\%$), for example, the average
path length is $\av{\ell}=13.4$, but only 4 of those links are needed to reach City~1. At this stage, Milgram's results seem quite within reach.

So far we have considered attrition only due to the strategy, or {\em strategical attrition}: the search is dropped  when the algorithm fails to find a next valid step.  In real life, however, there are other reasons for defecting besides the unavailability of an attractive option:  Participants may drop out from the experiment because of busyness, laziness, lack of motivation, etc.  We refer to this effect as {\em incidental attrition}.  We can lump both types of attrition into a single probability $p$ that an individual drops out of the experiment --- this means a path of length $\ell$ has $(1-p)^{\ell}$ chance of being completed.  From Milgram's second study~\cite{milgram69}, for example, it can be estimated that $p\approx0.38$.  To illustrate the effect of incidental attrition, in Fig.~\ref{fig_mixed_GP}(c) we plot the probability distribution  for paths of length $\ell$, $P(\ell)$, for the case of $\lambda=0.998$ (solid line), along with the distribution $(0.9)^{\ell}P(\ell)$ that results from an incidental dropout probability of $0.1$ (broken line).  As one would expect, the overall success rate drops, from $45\%$ to $11\%$,
but the (conditional) average path length is reduced by 4.3 links.  The two types of attrition are a significant factor in the selection of shorter paths.

\subsection{Asymmetry}
Consider finally the asymmetry of greedy, or decentralized paths: paths from $i$ in City~1 to $j$ in City~2 are not necessarily the same as paths from $j$ to $i$.  We see this effect quite clearly in Fig.~\ref{fig_migration_GP}, where
the average path length for City~$1\to2$ is {\em systematically} shorter than for City~$2\to1$, through all stages of the the model's buildup.  The success rates, too, are systematically smaller for paths from City~1 to~2 than the reverse (the differences are small and in the figure we put, for simplicity, only the average of the two rates).

A simple explanation to this asymmetry is that purely greedy paths from City~1 to City~2 can go through City~3, but those from City~2 to City~1 cannot (City~3 is {\em farther away} from the target), see Fig.~\ref{fig_mega}(b). The situation is statistically symmetric for a ``direct" commute, City~$1\leftrightarrow2$, without City~3 in the picture: same expected number of successful paths and average path lengths in either direction. The extra $2\to3\to1$ routes tend to be longer than the direct commute, and account both for the higher success rate and the longer average path lengths in the City~$2\to1$ direction.

We  observe small similar asymmetries also with our mixed strategy, for all values of $\lambda$.  The region where the mixed strategy beats the pure greedy algorithm, for example, is somewhat narrower for the City~$1\to2$ direction, with $\lambda_1=0.988$ (instead of $\lambda_1=0.986$ for City~$2\to1$), but we do not have a simple explanation to account for these findings.

\section{Discussion}
In summary, we have proposed improvements to the baseline SSCN model of~\cite{Frasco2014} that render it suitable for simulations of dynamic social processes, such as Milgram's Small-World experiment~\cite{milgram,milgram69}.  The most important revisions call for connecting each node to a number of spatially closest nearest neighbors, to account for ``next-door" friends, and relocating a fraction $\varepsilon$ of the nodes, to account for relocations (due to job change, study, marriage, etc.).  These two revisions have a minor effect on the degree distribution of the baseline model, but a dramatic effect on the connectivity properties of the network of social contacts:  The connections to closest neighbors make for a robust clustering effect (absent in the baseline model),
and even a tiny fraction $\varepsilon$ of relocations introduces long-range connections that decrease the average path length between pairs of nodes substantially, similarly to the random long-range links in Watts and Strogatz's Small-World networks~\cite{Watts98}.

Our simulations of the Milgram Small-World experiment show that Kleinberg's greedy algorithm --- based only on the geographical distance between nodes --- is successful in finding decentralized paths between pairs of nodes, but the paths are too long to explain Milgram's results.  We have shown that more complex strategies, such as occasionally  passing the message to acquaintances that are especially well-connected, can result in a significant reduction of the path length.  We have also confirmed the notion that geography is the most important consideration in finding short paths~\cite{dodds03,liben-nowell05}, at least
in the initial stages, until the message reaches the target's city.   The remaining path to the target, within the city, could be shortened considerably using the additional explicit information (e.g, occupation) and implicit information (ethnicity, social status) known about the target.  We have also discussed the effect of attrition (the fact that participants drop out of the experiment for various reasons) and showed how it helps select for shorter paths.   

Simulations of Milgram's experiment pose a particularly strict test to the SSCN model, in that finding decentralized paths relies quite sensitively both on the location of the nodes and on their network of connections.  The model's success makes it a promising substrate for the simulation of other dynamical processes on social networks, where such considerations are important (epidemics, opinion models, etc.).

\acknowledgements\vspace{-0.1in}
This work was funded in part by the Simons Foundation Grant No. 318812 and the Army Research Office Grant No. W911NF-16-1-0081.

\bibliographystyle{apsrev4-1}

\end{document}